\documentclass[onecolumn]{article}

\usepackage{fontenc}
\usepackage[latin1]{inputenc}
\usepackage[english]{babel}
\usepackage{graphicx}
\usepackage{amssymb}
\usepackage{amsmath}
\usepackage{lineno}

\begin{document}
\markboth{C. A. Duarte}{Considerations on the electromagnetic field wave equation and the Dirac equation}

\title{Considerations on the electromagnetic field wave equation and the Dirac equation}

\author{Celso de Araujo Duarte \\Departamento de F\'isica\\
Universidade Federal do Paran\'a\\
CP 19044 -- 81531-990\\
Curitiba (PR), Brazil\\
celso@fisica.ufpr.br}

\maketitle


\begin{abstract}
Here is constructed a heuristic, first-order differential equation for the electromagnetic field in the vacuum, based on a phenomenological \textsl{ad hoc} argument. The formal similarity between this \textsl{ad hoc} wave equation and the Dirac equation is explored.
\end{abstract}






\section{Introduction}\label{Intro}

It is a fact the universal acceptance of a second-order wave equation, to describe the propagation of the electromagnetic waves, arising from reasons of conceptual nature. However, we start this Introduction with some historical notes, where either one- or two-order wave equations were considered.

Historically, Riemann proposed \cite{whittaker} that the propagation of either electric, magnetic, optic and even gravitational disturbances should be the consequence of compression resistance properties of aether, proposing a generalization of the Poisson equation for static potentials $V$:
\begin{equation}
\nabla^2 V +\epsilon_0^{-1}{\rho}=0,
\end{equation}
(where $\rho$ is the associated charge density) to the most general time-dependent equation,
\begin{equation}\label{riemann}
\square V+\epsilon_0^{-1}{\rho}=0,
\end{equation}
as a result of his attempts to realize Gauss long ago idea, that the known forces which act between the charges would be supplemented by other forces, such as would cause the electric (magnetic, gravitational, etc.) actions to be propagated between the charges with a constant velocity \cite{whittaker}.

Soon after (1861-62), Maxwell obtained \cite{maxwell} the second-order wave equation (SOWE) for the magnetic field vector $\boldsymbol B$,
\begin{equation}\label{wave_eq}
\partial^2_t\boldsymbol B=c^2\nabla^2 \boldsymbol B \ \Longleftrightarrow \square \boldsymbol B = 0.
\end{equation}

In 1867, Ludvig Lorenz started from Neumann's theory \cite{whittaker} in which the electric field $\boldsymbol E$ was expressed by
\begin{equation}\label{neumann}
\boldsymbol E=\nabla \phi - c^{-1}\partial_t \boldsymbol A,
\end{equation}
where $\phi$ is the scalar electrostatic potential and $\boldsymbol A$ is the vector potential, and proposed the concept of retarded potentials \cite{lorenz} which required that
\begin{equation}\label{lorenz}
\square\phi = -\epsilon_0^{-1}\rho; \ \ \ \ \square\boldsymbol A = -\mu_0\boldsymbol J,
\end{equation}
where $\boldsymbol J$ is the current density vector. Writing in terms of the four-potential $\phi_{\mu}$ ($\phi_0=\phi$, $\phi_i=A_i$, $i=1,2,3$), it results the second order equation,
\begin{equation}\label{wave}
\partial^{\nu}\partial_{\nu}\phi_{\mu} = \square\phi_{\mu}= -\epsilon_0^{-1}j_{\mu},
\end{equation}
where $j_{\mu}$ is the four-current. Note that on the above works, the wave equation was written for either the potential or the field.

On the special relativity theory, the relativistically invariant four-vector potential $\phi^{\mu}$ plays the central role, while the electric $\boldsymbol E$ and the magnetic $\boldsymbol B$ fields are only three-dimensional vectors whose components appear sparsely in the components of the relativistically invariant electromagnetic tensor. The four-potential $\phi_{\mu}$ is also the central root in quantum electrodynamics (QED), instead of the physical fields $\boldsymbol E$ and $\boldsymbol B$, since the formalism of QED is based on the second quantization of fields, that starts with the Fourier series decomposition of the four-potential $\phi^{\mu}$ in terms of the nonclassical creation and annihilation operators.

However, as we will see on the following, the change of the hierarchy potentials \textit{vs.} fields reappears in papers on the 20$^{th}$ century. Another issue considered yet on the 20$^{th}$ century was the question about the convenience of a first order wave equation (FOWE) for the description of the electromagnetic waves, instead of the well-known SOWE forms (as equations \ref{wave_eq}, \ref{lorenz}, and \ref{wave}). Yet in 1931, Oppenheimer\cite{oppen} argued that equation \ref{wave} is in several respects unsatisfactory, and tried the development of a FOWE, reporting the previous equation constructed by Jordan\cite{jordan} using the Pauli matrices $\sigma_i$,
\begin{equation}\label{jordan_eq}
\left\{\left(\boldsymbol{\sigma}\cdot\nabla\right)+\partial_t\right\}\phi=0,
\end{equation}
however, $\phi$ would not be invariant under spatial rotations and by this reason should refer not to vectors (as required) but to spinors. Then, Oppenheimer presented the equation
\begin{equation}\label{oppen_eq}
\left\{\left(\boldsymbol{\tau}\cdot\nabla\right)+\partial_t\right\}\phi=0,
\end{equation}
using 3$\times$3 matrices $\tau_i$; and in order to satisfy Lorentz invariance, wrote
\begin{equation}\label{oppen_eq2}
\left\{\left(\boldsymbol{\rho}\cdot\nabla\right)+\partial_t\right\}\phi=0,
\end{equation}
defining 4$\times$4 matrices $\rho_i$, now with a four-component $\phi$, such that equation \ref{oppen_eq2} reduces to equation \ref{oppen_eq} setting equal to zero the fourth component, $\phi_4=0$. This leads to null three-dimension divergences, whose Lorentz invariance is gained (and also gained by equation \ref{oppen_eq}) transforming $\phi$ not being a four-vector, but as the components of a self dual six vector $\Phi_{\mu\nu}$, completely determined just by:
\begin{equation}
\Phi_{41}=i/\sqrt2(\psi_1+\psi_3); \ \ \ \Phi_{42}=i/\sqrt2(\psi_1-\psi_3); \ \ \ \Phi_{43}=-i\psi_2; \ \ \ \Phi_{44}=0,
\end{equation}

The solution of a linear equation provides directly the four-potential $\phi^{\mu}$ from a single set of boundary conditions (one set integration constants), while the traditional SOWE for the ER in the vacuum \ref{wave}, $\square\phi^{\mu}=0$, requires two sets of integration constants. In this sense, the traditional form $\square\phi^{\mu}=0$ can be considered in some sense ``redundant''.

On the counterpart, Mignani et al.\cite{mign} stressed the importance of the magnitude $\boldsymbol E-ic\boldsymbol B$, in the context of the electrodynamics, which is just the complex conjugate of well-known Riemann-Silberstein vector \cite{silb,birula} $\boldsymbol G=\boldsymbol E+ic\boldsymbol B$. Then, they appeal to an idea of Majorana of a Dirac-like equation for the photon \cite{majo}, expressing a probability quantum wave of the photon in terms of $\boldsymbol E$, $\boldsymbol B$ in a FOWE, where the fields are the relevant magnitudes instead of the potentials. In such a formalism, two of the Maxwell equations are condensed on the eigenvalue equation $\hat H \boldsymbol G=ic\hat{\boldsymbol p}\times\boldsymbol G$, with energy eigenvalues $E=cp$. This leads to the electromagnetic radiation (ER) energy density $\left|\boldsymbol G\right|^2\propto u$ equivalent to the photon probability density. In this formalism, the four potential $\phi^{\mu}$ has not central importance.

Such lines of reasoning were considered later by Giannetto \cite{gian}, who even generalized the idea of a FOWE to a non-Abelian charged field \cite{gian2} for potentials.

These preliminary considerations reveal: the attempts to construct a FOWE; and the relativism of the hierarchy of the potential with respect to the field, alongside diverse tentative models.

The present work follows a different strategy constructing an \textsl{ad hoc}, heuristic FOWE for the electromagnetic field, based on the phenomenology of the propagation of the ER, though the purpose is not to advocate the replacement of the SOWE by a FOWE. 

In the sequence of the article, this FOWE is compared to the Dirac equation. An algebraic manipulation involving the components of the Dirac spinor reveals close formal parallel with expressions of the theory of the ER.


\section{The heuristic ER wave equation}

In this section, we show the construction of a FOWE for the ER by very simple and phenomenological artifact, based on the spatial arrays of the electric and the magnetic fields with respect to the direction of propagation of the electromagnetic waves.

Let us consider the plane electromagnetic waves in the vacuum. Considering the orthogonality of the electric and the magnetic fields,
\begin{equation}\label{E.B}
\boldsymbol E.\boldsymbol B=0\hspace{10pt}\Leftrightarrow\hspace{10pt} E_xB_x+E_yB_y+E_zB_z=0.
\end{equation}
Note that $\boldsymbol E.\boldsymbol B$ is an invariant of the electromagnetism, and consequently relation \ref{E.B} is Lorentz invariant.

We will adopt a reference frame where the projections of $\boldsymbol E$ and $\boldsymbol B$ on the plane $xy$ are mutually perpendicular. On this frame,
\begin{equation}\label{ExBy}
E_x=B_y,\hspace{20pt}E_y=-B_x.
\end{equation}
An objection could be made with respect to this restriction. However, this does not prejudice the Lorentz invariance. In fact, once it is chosen a second reference frame $S'$ moving with velocity $v$ with respect to the given reference frame $S$, we may guide the orientation of the new axis $x',y',z'$ respectively parallel to $x,y,z$, to arrive at similar relations as \ref{ExBy} for the new components of the electric $\boldsymbol E'$ and the magnetic $\boldsymbol B'$ fields. Note that the arbitrary choice of the orientation of axis is neither incorrect, nor incompatible with the special relativity, nor a new issue, since it is commonly applied in the context of spin and angular momentum in quantum mechanics.

Expressions \ref{E.B} and \ref{ExBy} imply that
\begin{equation}\label{cond}
E_z=0 \hspace{10pt} \textrm{and} \hspace{10pt} B_z=0 \hspace{2pt}.
\end{equation}

Conditions \ref{ExBy} and \ref{cond} can be written explicitly as a function of the four-potential. Doing so and grouping the resulting expressions (from \ref{ExBy} and, for example, the first  of \ref{cond}), we obtain (adopting Einstein summation rule)
\begin{equation}\label{3equs}
\left.
\begin{array}{r}
-\partial_0\phi_1-\partial_1\phi_0-\epsilon_{2,jk}\partial_j\phi_k=0\\
-\partial_0\phi_2-\partial_2\phi_0+\epsilon_{1,jk}\partial_j\phi_k=0\\
-\partial_0\phi_3-\partial_3\phi_0=0\\
\end{array}
\right\}
\end{equation}

Now we also consider the Lorentz gauge $\partial_{\mu}\phi^{\mu}=0$, and group it to equations \ref{3equs}, to construct the matrix equation:
\begin{equation}\label{linear}
\varepsilon^{\mu}\partial_{\mu}\phi=0
\end{equation}
which is an heuristic FOWE for the ER ($\phi=(\phi^0,\phi^1,\phi^2,\phi^3)$), where we have defined the four 4$\times$4 matrices:
\begin{equation}\label{e}
\begin{array}{ll}
\varepsilon^0=\left(\begin{smallmatrix}
\ 0&-1\ &0&0\\ \ 0&0&-1\ &0\\ \ 0&0&0&-1\ \\ \ 0&0&0&0
\end{smallmatrix}\right), &
\varepsilon^1=\left(\begin{smallmatrix}
-1\ & \ 0& \ 0& \ 1 \ \\0& \ 0& \ 0& \ 0 \ \\0& \ 0& \ 0& \ 0 \ \\0& \ 1& \ 0& \ 0 \ 
\end{smallmatrix}\right),\\
\\
\varepsilon^2=\left(\begin{smallmatrix}
0& \ 0& \ 0& \ 0 \ \\-1\ & \ 0& \ 0& \ 1 \ \\0& \ 0& \ 0& \ 0 \ \\0& \ 0& \ 1& \ 0 \ 
\end{smallmatrix}\right), &
\varepsilon^3=\left(\begin{smallmatrix}
0&-1\ &0& \ 0 \ \\0&0&-1\ & \ 0 \ \\-1\ &0&0& \ 0 \ \\0&0&0& \ 1 \ 
\end{smallmatrix}\right).\\
\end{array}
\end{equation}
whose asymmetry arises from the chosen orientation of the reference frame. We stress the Lorentz invariance of equation \ref{linear}. Such an equation could be constructed in a different way (e.g. choosing the second expression of \ref{cond} instead of the first), leading to different matrices $\epsilon^{\mu}$, which is not relevant for our purposes. On the following, the most important result is that the ER can be described by a FOWE in terms of the components of $\phi$ as \ref{linear} (or as in the previous historical works mentioned on Section \ref{Intro}).


\section{The Dirac equation: a parallel}\label{para}

Consider the Dirac equation for massless fermions ($\hbar=1$),
\begin{equation}\label{dirac0}
i\gamma^{\mu}\partial_{\mu}\Psi=0.
\end{equation}
We are led to apply the inverse route of mathematical manipulation we followed to obtain equation \ref{linear}. For that, we write explicitly the spinor as $\Psi=\left(\psi_n+i\chi_n\right)$, where $\psi_n,\chi_n\in \Re, n=0,...3$ are respectively the real and the imaginary parts of each component of $\Psi$. By direct substitution on \ref{dirac0}, we obtain the following set of equations:
\begin{equation}\label{real}
\left(
\begin{array}{c}
\partial_2 \chi_3 + \partial_0 \psi_0 + \partial_3 \psi_2 + \partial_1 \psi_3\\
-\partial_2 \chi_2 + \partial_0 \psi_1 + \partial_1 \psi_2 - \partial_3 \psi_3\\
-\partial_2 \chi_1 - \partial_3 \psi_0 - \partial_1 \psi_1 - \partial_0 \psi_2\\
\partial_2 \chi_0 - \partial_1 \psi_0 + \partial_3 \psi_1 - \partial_0 \psi_3
\end{array}
\right)=0
\end{equation}
and
\begin{equation}\label{imag}
\left(
\begin{array}{c}
\partial_0 \chi_0 + \partial_3 \chi_2 + \partial_1 \chi_3 - \partial_2 \psi_3\\
\partial_0 \chi_1 + \partial_1 \chi_2 - \partial_3 \chi_3 + \partial_2 \psi_2\\
-\partial_3 \chi_0 - \partial_1 \chi_1 - \partial_0 \chi_2 + \partial_2 \psi_1\\
-\partial_1 \chi_0 + \partial_3 \chi_1 - \partial_0 \chi_3 - \partial_2 \psi_0
\end{array}
\right)=0
\end{equation}

Let us make the change of labeling: 
\begin{equation}\label{change}
\left(
\begin{array}{c} \psi_0\\ \psi_1\\ \psi_2\\ \psi_3\\ \end{array}
\right)
\mapsto
\left(
\begin{array}{r}\psi_0\\ -\chi_2\\ \psi_3\\ \psi_1\\ \end{array}
\right);  \hspace{5pt}
\left(
\begin{array}{c}\chi_0\\ \chi_1\\ \chi_2\\ \chi_3\\ \end{array}
\right)
\mapsto
\left(
\begin{array}{r}\chi_3\\ \chi_1\\ \chi_0\\ \psi_2\\ \end{array}
\right).
\end{equation}

We define the following three-dimensional, three-component magnitudes $\boldsymbol E_{\psi\left\{\chi\right\}}$ and $\boldsymbol B_{\psi\left\{\chi\right\}}$:
\begin{equation}\label{new}
\begin{array}{ll}
\boldsymbol E_{\psi}=-\partial_0\boldsymbol{\psi}-\nabla{\psi_0};\hspace{10pt}& \boldsymbol B_{\psi}=\nabla\times\boldsymbol{\psi};\\
\boldsymbol E_{\chi}=-\partial_0\boldsymbol{\chi}-\nabla{\chi_0};\hspace{10pt}& \boldsymbol B_{\chi}=\nabla\times\boldsymbol{\chi},\\
\end{array}
\end{equation}
which satisfies the same algebra of the usual electromagnetic fields $\boldsymbol E$ and $\boldsymbol B$ with respect to the four-vector electromagnetic potential $\phi^{\mu}$. Despite until now without physical meaning, these magnitudes $\boldsymbol E_{\psi\left\{\chi\right\}}$ and $\boldsymbol B_{\psi\left\{\chi\right\}}$ have components on the three dimensional space as well as $\boldsymbol{\psi}$ and $\boldsymbol{\chi}$. However, $\boldsymbol E_{\psi\left\{\chi\right\}}$ and $\boldsymbol B_{\psi\left\{\chi\right\}}$ are not vectors, since $\boldsymbol{\psi}$ and $\boldsymbol{\chi}$ do not transform as genuine vectors. Though $\boldsymbol E_{\psi\left\{\chi\right\}}$ and $\boldsymbol B_{\psi\left\{\chi\right\}}$ are fields in the space, they are not of vector nature.

The substitution of these magnitudes in expressions \ref{real} and \ref{imag} gives:
\begin{equation}\label{final}
\left(
\begin{array}{c}
\partial_{\mu}\psi^{\mu}\\ 
E_{\chi,2} -B_{\psi,2}\\
E_{\psi,3} +B_{\chi,3}\\
E_{\psi,1} +B_{\chi,1}\\
\end{array}
\right)=0\hspace{10pt}\textrm{and}\hspace{10pt}
\left(
\begin{array}{c}
E_{\chi,3} -B_{\psi,3}\\
E_{\chi,1} -B_{\psi,1}\\
\partial_{\mu}\chi^{\mu}\\
E_{\psi,2} +B_{\chi,2}\\
\end{array}
\right)=0,
\end{equation}
which can be writen as
\begin{equation}\label{rel}
\boldsymbol E_{\psi}=-\boldsymbol B_{\chi}\hspace{15pt} \textrm{and} \hspace{15pt} \boldsymbol E_{\chi}=\boldsymbol B_{\psi},
\end{equation}
together with the conditions
\begin{equation}\label{gauge}
 \partial_{\mu}\chi^{\mu}=0\hspace{15pt} \textrm{and} \hspace{15pt} \partial_{\mu}\psi^{\mu}=0.
\end{equation}
As can be seen, relations \ref{rel} between the fields $\boldsymbol E_{\psi\left\{\chi\right\}},\boldsymbol B_{\psi\left\{\chi\right\}}$ are the equivalent, for  the massless fermion, to the photon relations \ref{ExBy}, and relations \ref{gauge} are ``gauge like'' conditions (analogous to the Lorentz gauge condition $\partial_{\mu}\phi^{\mu}$). As a partial conclusion, the spinor field $\Psi$ for massless fermions leaves to the set of (nonvector) fields $\boldsymbol E_{\psi\left\{\chi\right\}},\boldsymbol B_{\psi\left\{\chi\right\}}$ of the three dimensional space.

In particular, for other alternative permuted relabeling (instead of \ref{change}), we arrive to the same results, apart from irrelevant changes of signal.
\newline

Now we treat the most general case of massive fermions, for which Dirac equation \ref{dirac0} has the mass term:
\begin{equation}\label{diracm}
i\gamma^{\mu}\partial_{\mu}\Psi=m\Psi.
\end{equation}

Explicitly,
\begin{equation}\label{mRe}
\left(
\begin{array}{c}
\partial_{\mu}\psi^{\mu}\\
E_{\chi,2} -B_{\psi,2}\\
E_{\psi,3} +B_{\chi,3}\\
E_{\psi,1} +B_{\chi,1}\\
\end{array}
\right)=
m\left(
\begin{array}{r}
\chi_3\\
\chi_1\\
\chi_0\\
\psi_2\\
\end{array}
\right)
\end{equation}
and
\begin{equation}\label{mIm}
\left(
\begin{array}{c}
E_{\chi,3} -B_{\psi,3}\\
E_{\chi,1} -B_{\psi,1}\\
\partial_{\mu}\chi^{\mu}\\
E_{\psi,2} +B_{\chi,2}\\
\end{array}
\right)=
m\left(
\begin{array}{r}
\psi_0\\
-\chi_2\\
\psi_3\\
-\psi_1\\
\end{array}
\right),
\end{equation}
which were obtained respectively from the real and the imaginary parts of the Dirac massive fermion equation after the application of the relabeling \ref{change}. It is interesting to consider the components of the four-vector fermion current density $j^{\mu}=\overline{\psi}\gamma^{\mu}\gamma^0\psi$. It can be written as
\begin{equation}\label{j}
\left.
\begin{array}{lll}
j^0&=&m^{-2}\left(\boldsymbol C^2+\boldsymbol D^2\right)+\psi_3^2+\chi_3^2\\
\boldsymbol j&=&2m^{-2}\left(\boldsymbol C\times\boldsymbol D\right)+\boldsymbol{\Lambda}\\
\end{array}
\right\}.
\end{equation}
The expressions in \ref{j} were written in a compact form in terms of the three-dimensional field $\boldsymbol{\Lambda}$ with the components
\begin{equation}\label{lambda}
\left.
\begin{array}{lll}
\Lambda_{1,2}&=&-\left(\boldsymbol{\psi}\times\boldsymbol{\chi}\right)_{1,2}\\
 \Lambda_3&=&(\chi_0\chi_3+\psi_0\psi_3)
\end{array}
\right\},
\end{equation}
and of the three-dimensional fields,
\begin{equation}\label{CD}
\begin{array}{ll}
\boldsymbol C=\boldsymbol E_{\chi}-\boldsymbol B_{\psi};\hspace{10pt}&\boldsymbol D=\boldsymbol E_{\psi}+\boldsymbol B_{\chi}.\\
\end{array}
\end{equation}

Note that on the particular case of massless fermions, $\boldsymbol C$ and $\boldsymbol D$ vanish, resulting the equalities \ref{rel}, and also in this case, $j^0\equiv\psi_3^2+\chi_3^2$ and $\boldsymbol j\equiv\boldsymbol{\Lambda}$.

Expressions \ref{j} are strikingly similar to, respectively, the expressions of the electromagnetic energy density and the Poynting vector,
\begin{equation}\label{u}
\left.
\begin{array}{l}
u=(2\varepsilon_0)^{-1}(\boldsymbol E^2+\boldsymbol B^2)\\
\boldsymbol S=\varepsilon_0^{-1}(\boldsymbol E\times\boldsymbol B)
\end{array}
\right\},
\end{equation}
which can be seen with the substitutions $\boldsymbol C\longmapsto \boldsymbol E$, $\boldsymbol D \longmapsto \boldsymbol B$ and $m^2 \longmapsto 2\varepsilon_0$, apart from the new additive terms $\psi_3^2+\chi_3^2$ and $\boldsymbol \Lambda$, which have not any corresponding similar on the electromagnetic theory. The similarity between the magnitude $j^0$ in expression \ref{j} and the energy density \textit{u} in expression \ref{u} may be considered more than a simple mathematical similarity, since the time flux $j^0$ represents, apart from a multiplicative factor, the energy flux associated strictly with the fermion rest mass \textit{m} (that corresponds to the total amount of energy $mc^2=m$, in our system of units).

The parallel presented above between the FOWE for the ER and the Dirac equation left to the emergence of the fields $\boldsymbol C,\boldsymbol D$ (of nonvector nature). The mathematical similarity of the expressions of the current and energy densities, eqs. \ref{j} with their correspondents for the ER, leaves to consider that, while the electric and the magnetic fields $\boldsymbol E$ and $\boldsymbol B$ are real physical magnitudes that can be measured through their actions on electrically charged particles, the fields $\boldsymbol E_{\psi\left\{\chi\right\}},\boldsymbol B_{\psi\left\{\chi\right\}}$ or either $\boldsymbol C,\boldsymbol D$ may also correspond to physical interaction and be experimentally detected as the signature of fermions. If this is valid, while the electric and the magnetic fields satisfy globally the Maxwell equations, we can speculate if equations \ref{rel} or \ref{CD} -- and consequently the Dirac equation itself -- corresponds to a ``fermionic radiation field'' that is globally governed by a particular set of field equations. In fact, from \ref{new} we obtain
\begin{equation}
\begin{array}{ll}
\nabla\times\boldsymbol E_{\psi}=-\partial_0\boldsymbol B_{\psi};\hspace{10pt}& \nabla\times\boldsymbol B_{\psi}=\nabla\pi-\nabla^2\boldsymbol{\psi};\\
\nabla\times\boldsymbol E_{\chi}=-\partial_0\boldsymbol B_{\chi};\hspace{10pt}& \nabla\times\boldsymbol B_{\chi}=\nabla\kappa-\nabla^2\boldsymbol{\chi};\\
\\
\nabla\cdot\boldsymbol E_{\psi}=-\partial_0\pi-\nabla^2{\psi_0};\hspace{10pt}& \nabla\cdot\boldsymbol B_{\psi}=0;\\
\nabla\cdot\boldsymbol E_{\chi}=-\partial_0\kappa-\nabla^2{\chi_0};\hspace{10pt}& \nabla\cdot\boldsymbol B_{\chi}=0;\\
\end{array}
\end{equation}
which are the equivalent to the Maxwell equations, with $\pi=\nabla\cdot\boldsymbol{\psi}$, and $\kappa=\nabla\cdot\boldsymbol{\chi}$. 

It can be argued that there is not any experimental evidence to justify the distinction of these fields in nature, and the present work should be considered only of formal value. In the other case, it could be wondered if the fields $\psi,\chi$ are associated with the yet unidentified way to a physical interaction. 



\end{document}